\documentclass[a4paper,aps,twocolumn,nofootinbib]{revtex4}
\RequirePackage[colorlinks,hyperindex]{hyperref}
\RequirePackage[english]{babel}
\RequirePackage[latin1]{inputenc}
\RequirePackage[T1]{fontenc}
\RequirePackage{mathrsfs}
\RequirePackage{amsmath}
\RequirePackage{amssymb}
\RequirePackage{amsbsy}
\RequirePackage{color}
\RequirePackage{bm}
\hypersetup{colorlinks=true,breaklinks=true,urlcolor=blue,linkcolor=red}
\begin{document}
%%%%%%%%%%%%%%%%%%%%%%%%%%%%%%%%%%%%%%%%%%%%%%%%%%%%%%%%%%%%%%%%%%%%%%%%%%%%%%%%%%%%%%%%%%%%%%%%%%%
\title{\bf{Non-existence of rest-frame spin-eigenstate spinors in their own electrodynamics}}
\author{Luca Fabbri$^{\ast}$, Rold\~{a}o da Rocha$^{\dagger}$}
\affiliation{$^{\ast}$ DIME, Universit\`{a} di Genova,\\ P.Kennedy Pad.D, 16129 Genova, ITALY,\\
$^{\dagger}$ CMCC, Universidade Federal do ABC,\\ 09210-580, Santo Andr\'e, BRAZIL}
\date{\today}
%%%%%%%%%%%%%%%%%%%%%%%%%%%%%%%%%%%%%%%%%%%%%%%%%%%%%%%%%%%%%%%%%%%%%%%%%%%%%%%%%%%%%%%%%%%%%%%%%%%
\begin{abstract}
We assume a physical situation where gravity with torsion is neglected for an electrodynamically self-interacting spinor that will be taken in its rest-frame and spin-eigenstate: we demonstrate that under this circumstance no solution exists for the system of field equations. Despite such a situation might look artificial nevertheless it represents the instance that is commonly taken as the basis for all computations of quantum electrodynamics.
\end{abstract}
%%%%%%%%%%%%%%%%%%%%%%%%%%%%%%%%%%%%%%%%%%%%%%%%%%%%%%%%%%%%%%%%%%%%%%%%%%%%%%%%%%%%%%%%%%%%%%%%%%%
\maketitle
%%%%%%%%%%%%%%%%%%%%%%%%%%%%%%%%%%%%%%%%%%%%%%%%%%%%%%%%%%%%%%%%%%%%%%%%%%%%%%%%%%%%%%%%%%%%%%%%%%%
\section{Introduction}
Quantum electrodynamics (QED) is one of the most successful models ever developed in physics, although the need of renormalizability, ultra-violet divergences and the non-convergence of perturbative series make it clear that a proper systematization is needed. While it is possible to think that for such a predictive set of rules most of the problems may be solved in terms of rigorous mathematical treatments, results like the Haag theorem \cite{Haag:1955ev,s-w} seem to inexorably undermine the essence of field quantization.

In this paper however, we are not going to consider the protocols of field quantization at all. In fact, we shall be focusing on a purely classical theory of electrodynamics, but we will see there are still consistency problems under the assumptions and hypotheses that are usually made.

These assumptions and hypotheses consist in neglecting torsion with gravity and in postulating that it is always possible to find a system of reference where a single fermion can be taken at rest and with spin aligned along the third axis: as a matter of fact, all this appears to be very reasonable, but despite of this, precise mathematical implementations of these ideas will lead toward a system of field equations that can be proven to have no solution.
%%%%%%%%%%%%%%%%%%%%%%%%%%%%%%%%%%%%%%%%%%%%%%%%%%%%%%%%%%%%%%%%%%%%%%%%%%%%%%%%%%%%%%%%%%%%%%%%%%%
%%%%%%%%%%%%%%%%%%%%%%%%%%%%%%%%%%%%%%%%%%%%%%%%%%%%%%%%%%%%%%%%%%%%%%%%%%%%%%%%%%%%%%%%%%%%%%%%%%%
\section{Geometry of Dirac Spinor Fields}
We begin by recalling that $\boldsymbol{\gamma}^{a}$ are matrices belonging to the Clifford algebra, from which $[\boldsymbol{\gamma}^{a}\!,\!\boldsymbol{\gamma}^{b}]\!=\! 4\boldsymbol{\sigma}^{ab}$ defines the set of generators while $2i\boldsymbol{\sigma}_{ab}\!=\!\varepsilon_{abcd}\boldsymbol{\pi}\boldsymbol{\sigma}^{cd}$ defines the parity-odd $\boldsymbol{\pi}$ matrix\footnote{This matrix is what is usually indicated as gamma with an index five, but since in the space-time this index has no meaning we prefer to use a notation with no index at all.} in a covariant and implicit manner.

As it is well known, Dirac spinor fields can be classified in terms of the so-called Lounesto classification according to two classes: \emph{singular} spinor fields are those subject to the conditions $i\overline{\psi}\boldsymbol{\pi}\psi\!\equiv\!0$ and $\overline{\psi}\psi\!\equiv\!0$ while \emph{regular} spinor fields are all those for which the two above conditions do not identically hold \cite{L,Cavalcanti:2014wia,Fabbri:2016msm}. For regular spinor fields, it is always possible to perform what is known to be the polar decomposition of the Dirac spinor field given in the form
\begin{eqnarray}
&\!\psi\!=\!\phi\sqrt{\frac{2}{\gamma+1}}e^{-i\alpha}e^{-\frac{i}{2}\beta\boldsymbol{\pi}}
\!\left(\!\begin{tabular}{c}
$\left(\frac{\gamma+1}{2}\boldsymbol{\mathbb{I}}\!-\!
\gamma\vec{v}\!\cdot\!\vec{\frac{\boldsymbol{\sigma}}{2}}\right)\xi$\\
$\left(\frac{\gamma+1}{2}\boldsymbol{\mathbb{I}}\!+\!
\gamma\vec{v}\!\cdot\!\vec{\frac{\boldsymbol{\sigma}}{2}}\right)\xi$
\end{tabular}\!\right)
\label{spinor}
\end{eqnarray}
up to the $\psi'\!=\!\boldsymbol{\pi}\psi$ transformation, with $\gamma\!=\!1/\!\sqrt{1\!-\!v^{2}}$ the relativistic factor, while $\xi$ such that $\xi^{\dagger}\xi\!=\!1$ is any generic $2$-component spinor, and the real $\alpha$ a phase parameter.

The bi-linear spinor fields are the di-pole moment
\begin{eqnarray}
&M_{ab}\!=\!2i\overline{\psi}\boldsymbol{\sigma}_{ab}\psi
\!=\!2\phi^{2}(u^{j}s^{k}\varepsilon_{jkab}\cos{\beta}\!+\!u_{[a}s_{b]}\sin{\beta})
\label{M}
\end{eqnarray}
the axial-vector and vector
\begin{eqnarray}
&S^{a}\!=\!\overline{\psi}\boldsymbol{\gamma}^{a}\boldsymbol{\pi}\psi\!=\!2\phi^{2}s^{a}\label{S}\\
&U^{a}\!=\!\overline{\psi}\boldsymbol{\gamma}^{a}\psi\!=\!2\phi^{2}u^{a}\label{U}
\end{eqnarray}
and the pseudo-scalar and scalar
\begin{eqnarray}
&\Theta\!=\!i\overline{\psi}\boldsymbol{\pi}\psi\!=\!2\phi^{2}\sin{\beta}\label{b2}\\
&\Phi\!=\!\overline{\psi}\psi\!=\!2\phi^{2}\cos{\beta}\label{b1}
\end{eqnarray}
where $\phi$ and $\beta$ are the module and the Yvon-Takabayashi angle: one can easily prove that the directions are
\begin{eqnarray}
&s^{a}\!=\!\frac{1}{\gamma+1}\left(\!\begin{tabular}{c}
$2\gamma(\gamma\!+\!1)(\vec{v}\!\cdot\!\vec{\varsigma})$\\
$2(\gamma+1)\vec{\varsigma}\!+\!2\gamma^{2}(\vec{v}\!\cdot\!\vec{\varsigma})\vec{v}$
\end{tabular}\!\right)\label{s}\\
&u^{a}\!=\!\frac{1}{\gamma+1}\left(\!\begin{tabular}{c}
$\frac{1}{2}(\gamma\!+\!1)^{2}+\frac{1}{2}\gamma^{2}(\vec{v}\!\cdot\!\vec{v})$\\
$\gamma(\gamma\!+\!1)\vec{v}$
\end{tabular}\!\right)\label{u}
\end{eqnarray}
such that conditions $u_{a}u^{a}\!=\!-s_{a}s^{a}\!=\!1$ and $u_{a}s^{a}\!=\!0$ hold, with $\xi^{\dagger}\vec{\boldsymbol{\sigma}}\xi\!=\!2\vec{\varsigma}$ as the spin and $\vec{v}$ as the velocity, and with the module and YT angle as the only two true degrees of freedom that the spinor field can have in general.

Calling $\Omega^{a}_{b\pi}$ the spin connection, it is possible to define
\begin{eqnarray}
&\boldsymbol{\Omega}_{\mu}
=\frac{1}{2}\Omega^{ab}_{\phantom{ab}\mu}\boldsymbol{\sigma}_{ab}
\!+\!iqA_{\mu}\boldsymbol{\mathbb{I}}\label{spinorialconnection}
\end{eqnarray}
in terms of the spin connection and the gauge potential of charge $q$ and called spinorial connection. Remark that for the moment everything is in the torsionless case.

With the spinorial covariant derivative of (\ref{spinor}) we get
\begin{eqnarray}
\nonumber
&\boldsymbol{\nabla}_{\mu}\psi\!=\![\nabla_{\mu}\ln{\phi}\mathbb{I}
\!-\!\frac{i}{2}\nabla_{\mu}\beta\boldsymbol{\pi}+\\
&+i(qA_{\mu}\!-\!\nabla_{\mu}\alpha)\mathbb{I}
\!+\!\frac{1}{2}(\Omega_{ij\mu}\!-\!\partial_{\mu}\theta_{ij})\boldsymbol{\sigma}^{ij}]\psi
\label{decspinder}
\end{eqnarray}
from which 
\begin{eqnarray}
&\nabla_{\mu}s_{\nu}\!=\!(\partial\theta\!-\!\Omega)_{\rho\nu\mu}s^{\rho}\label{ds}\\
&\nabla_{\mu}u_{\nu}\!=\!(\partial\theta\!-\!\Omega)_{\rho\nu\mu}u^{\rho}\label{du}
\end{eqnarray}
where $\alpha$ is the phase parameter and $\theta_{ij}$ encode the three rapidities expressed through the velocity $\vec{v}$ as well as the three rotations expressed as coefficients of the semi-spinor $\xi$ such that $\xi^{\dagger}\xi\!=\!1$ in the most general circumstance.

For the dynamics, we assume the action given by
\begin{eqnarray}
\nonumber
&\mathscr{L}\!=\!\frac{1}{4}(\partial W)^{2}\!-\!\frac{1}{2}M^{2}W^{2}
\!+\!R\!+\!\frac{1}{4}F^{2}-\\
&-i\overline{\psi}\boldsymbol{\gamma}^{\mu}\boldsymbol{\nabla}_{\mu}\psi
\!+\!XS^{\mu}W_{\mu}\!+\!m\Phi
\label{l}
\end{eqnarray}
with $(\partial W)_{\mu\nu}$ the curl of $W_{\mu}$ being the torsion axial vector while $R$ is the Ricci scalar and $F_{\mu\nu}$ is the Faraday tensor, where $X$ is the strength of the interaction between torsion and the spin of spinor fields while $M$ and $m$ are the mass of torsion and the spinor field itself, respectively. Having defined the connection in the torsionless case it may seem we have restricted generality, but in reality we can still be in the most general situation even though the connection is torsionless so long as torsion is later included in the form of some supplementary massive axial vector field.

Varying the above Lagrangian functional with respect to the spinor field gives the Dirac spinor field equations
\begin{eqnarray}
&i\boldsymbol{\gamma}^{\mu}\boldsymbol{\nabla}_{\mu}\psi
\!-\!XW_{\mu}\boldsymbol{\gamma}^{\mu}\boldsymbol{\pi}\psi\!-\!m\psi\!=\!0\label{D}
\end{eqnarray}
and it is possible to demonstrate that with the polar form of the spinor there is a corresponding polar decomposition of the spinor field equations into the equivalent
\begin{eqnarray}
\nonumber
&\frac{1}{2}\varepsilon_{\mu\alpha\nu\iota}(\partial\theta\!-\!\Omega)^{\alpha\nu\iota}
\!-\!2(\nabla\alpha\!-\!qA)^{\iota}u_{[\iota}s_{\mu]}+\\
&+2(\nabla\beta/2\!-\!XW)_{\mu}\!+\!2s_{\mu}m\cos{\beta}\!=\!0\label{f1}\\
\nonumber
&(\partial\theta\!-\!\Omega)_{\mu a}^{\phantom{\mu a}a}
\!-\!2(\nabla\alpha\!-\!qA)^{\rho}u^{\nu}s^{\alpha}\varepsilon_{\mu\rho\nu\alpha}+\\
&+2s_{\mu}m\sin{\beta}\!+\!\nabla_{\mu}\ln{\phi^{2}}\!=\!0\label{f2}
\end{eqnarray}
as field equations specifying all first-order derivatives of the module and the YT angle: the polar decomposition of the spinor field equations is essentially the way in which the four complex spinor field equations can be converted into two real vector field equations, which therefore turn out to be much easier to manipulate \cite{h1,Fabbri:2016laz}.

The geometric field equations are given by
\begin{eqnarray}
&\nabla_{\sigma}F^{\sigma\mu}\!=\!2q\phi^{2}u^{\mu}\label{m}
\end{eqnarray}
alongside to
\begin{eqnarray}
\nonumber
&R^{\rho\sigma}\!-\!\frac{1}{2}Rg^{\rho\sigma}
\!=\!\frac{1}{2}[M^{2}(W^{\rho}W^{\sigma}\!\!-\!\!\frac{1}{2}W^{\alpha}W_{\alpha}g^{\rho\sigma})+\\
\nonumber
&+\frac{1}{4}(\partial W)^{2}g^{\rho\sigma}
\!-\!(\partial W)^{\sigma\alpha}(\partial W)^{\rho}_{\phantom{\rho}\alpha}+\\
\nonumber
&+\frac{1}{4}F^{2}g^{\rho\sigma}\!-\!F^{\rho\alpha}\!F^{\sigma}_{\phantom{\sigma}\alpha}-\\
\nonumber
&-\phi^{2}[(XW\!-\!\nabla\frac{\beta}{2})^{\sigma}s^{\rho}
\!+\!(XW\!-\!\nabla\frac{\beta}{2})^{\rho}s^{\sigma}+\\
\nonumber
&+(qA\!-\!\nabla\alpha)^{\sigma}u^{\rho}\!+\!(qA\!-\!\nabla\alpha)^{\rho}u^{\sigma}-\\
&-\frac{1}{4}(\Omega\!-\!\partial\theta)_{ij}^{\phantom{ij}\sigma}\varepsilon^{\rho ijk}s_{k}
\!-\!\frac{1}{4}(\Omega\!-\!\partial\theta)_{ij}^{\phantom{ij}\rho}\varepsilon^{\sigma ijk}s_{k}]]
\label{e}
\end{eqnarray}
with
\begin{eqnarray}
&\nabla_{\alpha}(\partial W)^{\alpha\mu}\!+\!M^{2}W^{\mu}\!=\!2X\phi^{2}s^{\mu}\label{t}
\end{eqnarray}
as the field equations coupling electrodynamics, gravity and torsion to the currents, the energy and the spin \cite{Fabbri:2017fac}.
%%%%%%%%%%%%%%%%%%%%%%%%%%%%%%%%%%%%%%%%%%%%%%%%%%%%%%%%%%%%%%%%%%%%%%%%%%%%%%%%%%%%%%%%%%%%%%%%%%%
%%%%%%%%%%%%%%%%%%%%%%%%%%%%%%%%%%%%%%%%%%%%%%%%%%%%%%%%%%%%%%%%%%%%%%%%%%%%%%%%%%%%%%%%%%%%%%%%%%%
\section{Torsion-Electrodynamics for Spinor Fields}
So far everything is in its most general case.

Nevertheless, this very general theoretical situation is intractable in practice: the problem is the presence of the gravitational field, for the reason that follows.

Considering the spinor field equations in the polar form given by (\ref{f1}, \ref{f2}) it is clear that of the three irreducible parts of the spin connection, only the completely antisymmetric part and the trace part enter in the dynamics of the spinor field: this means that we can not even as a matter of principle write the gravitational information in terms of the spinorial degrees of freedom. Therefore, we must solve Einstein equations (\ref{e}) exactly for the spin connection, and this is an extremely difficult enterprise.

Some simpler situation can be encountered when gravity is neglected. Neglecting gravity would amount to have the constraint $\partial\theta\!-\!\Omega\!\equiv\!0$ and as a consequence a first advantage is that (\ref{ds}, \ref{du}) reduce to $\nabla_{\mu}s_{\alpha}\!=\!\nabla_{\mu}u_{\alpha}\!=\!0$ in all remaining equations: then (\ref{m}, \ref{t}) are always
\begin{eqnarray}
&\nabla_{\sigma}F^{\sigma\mu}\!=\!2q\phi^{2}u^{\mu}
\end{eqnarray}
and
\begin{eqnarray}
&\nabla_{\alpha}(\partial W)^{\alpha\mu}\!+\!M^{2}W^{\mu}\!=\!2X\phi^{2}s^{\mu}
\end{eqnarray}
but (\ref{f1}, \ref{f2}) become
\begin{eqnarray}
&(\frac{1}{2}\nabla\beta\!-\!XW)_{\mu}\!-\!(P\!-\!qA)^{\iota}u_{[\iota}s_{\mu]}
\!+\!s_{\mu}m\cos{\beta}\!=\!0\label{F1}\\
&\nabla_{\mu}\ln{\phi}\!-\!(P\!-\!qA)^{\rho}u^{\nu}s^{\alpha}\varepsilon_{\mu\rho\nu\alpha}
\!+\!s_{\mu}m\sin{\beta}\!=\!0\label{F2}
\end{eqnarray}
having called $\nabla\alpha\!=\!P$ for the sake of simplicity \cite{Fabbri:2017pwp}.

We remark that now both torsion and gauge potential are fully present in these field equations: on the one hand, we can invert (\ref{F1}) in order to make torsion explicit as  
\begin{eqnarray}
&XW_{\mu}\!=\!\frac{1}{2}\nabla_{\mu}\beta
\!-\!(P\!-\!qA)^{\iota}u_{[\iota}s_{\mu]}\!+\!s_{\mu}m\cos{\beta}\label{torsion}
\end{eqnarray}
while on the other hand, combining (\ref{F1}, \ref{F2}) gives
\begin{eqnarray}
\nonumber
&(P-qA)^{\nu}\!=\!m\cos{\beta}u^{\nu}
\!+\!s^{[\nu}u^{\mu]}(\frac{1}{2}\nabla\beta\!-\!XW)_{\mu}+\\
&+\varepsilon^{\nu\rho\sigma\mu}s_{\rho}u_{\sigma}\nabla_{\mu}\ln{\phi}\label{momentum}
\end{eqnarray}
showing that both torsion and gauge potential can in fact be written in terms of the spinorial degrees of freedom.

Then we can have torsion and gauge potentials substituted in terms of (\ref{torsion}, \ref{momentum}) into the torsional and electrodynamic field equations. Nevertheless, in (\ref{torsion}) we observe the presence of the gauge potential while in (\ref{momentum}) we see the presence of torsion, and as a consequence we cannot perform both substitutions at once removing both torsion and gauge potentials in both equations simultaneously.

In \cite{Fabbri:2017xlx} we studied the case where electrodynamics was absent to appreciate the effects of torsion, while here we neglect torsion to study electrodynamics for spinors.
%%%%%%%%%%%%%%%%%%%%%%%%%%%%%%%%%%%%%%%%%%%%%%%%%%%%%%%%%%%%%%%%%%%%%%%%%%%%%%%%%%%%%%%%%%%%%%%%%%%
%%%%%%%%%%%%%%%%%%%%%%%%%%%%%%%%%%%%%%%%%%%%%%%%%%%%%%%%%%%%%%%%%%%%%%%%%%%%%%%%%%%%%%%%%%%%%%%%%%%
\section{Electrodynamics with Spinors}
So as we have anticipated, here we will proceed to assume that no torsion is allowed in the field equations.

In the case of no torsion the spinor field equations are
\begin{eqnarray}
&\frac{1}{2}\nabla_{\mu}\beta\!-\!(P\!-\!qA)^{\iota}u_{[\iota}s_{\mu]}
\!+\!s_{\mu}m\cos{\beta}\!=\!0\label{1}\\
&\nabla_{\mu}\ln{\phi}\!-\!(P\!-\!qA)^{\rho}u^{\nu}s^{\alpha}\varepsilon_{\mu\rho\nu\alpha}
\!+\!s_{\mu}m\sin{\beta}\!=\!0\label{2}
\end{eqnarray}
developing the expression
\begin{eqnarray}
\nonumber
&(P-qA)^{\nu}\!=\!m\cos{\beta}u^{\nu}\!+\!s^{[\nu}u^{\mu]}\nabla_{\mu}\beta/2+\\
&+\varepsilon^{\nu\rho\sigma\mu}s_{\rho}u_{\sigma}\nabla_{\mu}\ln{\phi}\label{gauge}
\end{eqnarray}
which we plan to substitute into
\begin{eqnarray}
&\nabla_{\sigma}F^{\sigma\mu}\!=\!2q\phi^{2}u^{\mu}
\end{eqnarray}
in order to look for exact solutions.

When the expression of the gauge potential is inserted into the electrodynamic field equations we obtain 
\begin{eqnarray}
\nonumber
&2m\cos{\beta}u^{[\nu}\nabla^{\alpha]}\beta\nabla_{\alpha}\beta
\!+\!2m\sin{\beta}u^{[\nu}\nabla^{\alpha]}\nabla_{\alpha}\beta+\\
\nonumber
&+u^{\nu}s\!\cdot\!\nabla\nabla^{2}\beta\!-\!s^{\nu}u\!\cdot\!\nabla\nabla^{2}\beta+\\
&+s_{\rho}u_{\sigma}\varepsilon^{\mu\rho\sigma\nu}\nabla_{\mu}\nabla^{2}\ln{|2q\phi|^{2}}
\!=\!|2q\phi|^{2}u^{\nu}\label{ed}
\end{eqnarray}
while inserting it back into the spinor field equations gives
\begin{eqnarray}
&u_{\mu}u\!\cdot\!\nabla\ln{\phi}\!-\!s_{\mu}s\!\cdot\!\nabla\ln{\phi}
\!+\!s_{\mu}m\sin{\beta}\!=\!0\label{div}\\
&\nabla_{\mu}\beta\!-\!u_{\mu}u\!\cdot\!\nabla\beta
\!+\!s_{\mu}s\!\cdot\!\nabla\beta\!=\!0\label{constr}
\end{eqnarray}
as field equations in terms of the spinor field alone.

In the scalar product with $u$ and $s$ we obtain that (\ref{constr}) does not produce any relation while (\ref{div}) gives that
\begin{eqnarray}
&s\!\cdot\!\nabla\ln{\phi}\!-\!m\sin{\beta}\!=\!0\\
&u\!\cdot\!\nabla\ln{\phi}\!=\!0
\end{eqnarray}
whereas (\ref{ed}) gives
\begin{eqnarray}
\nonumber
&\nabla^{2}(u\!\cdot\!\nabla\beta)\!-\!2m\cos{\beta}s\!\cdot\!\nabla\beta u\!\cdot\!\nabla\beta-\\
&-2m\sin{\beta}s\!\cdot\!\nabla(u\!\cdot\!\nabla\beta)\!=\!0\\
\nonumber
&\nabla^{2}(s\!\cdot\!\nabla\beta)\!-\!2m\cos{\beta}|s\!\cdot\!\nabla\beta|^{2}-\\
&-2m\sin{\beta}s\!\cdot\!\nabla (s\!\cdot\!\nabla\beta)\!=\!|2q\phi|^{2}
\end{eqnarray}
where in the last ones we used (\ref{constr}) to simplify: plugging them back into the original field equations gives that (\ref{div}) is verified while (\ref{ed}) simplifies down to the form
\begin{eqnarray}
&s_{\rho}u_{\sigma}\varepsilon^{\mu\rho\sigma\nu}\nabla_{\mu}\nabla^{2}\ln{|2q\phi|^{2}}\!=\!0
\end{eqnarray}
which are divergenceless as it is to be expected from the fact that the conservation of the electric charge is already granted by the validity of the spinor field equations.

Altogether, the entire system of field equations consists of the following set of constraints
\begin{eqnarray}
&\nabla_{\mu}\beta\!-\!u_{\mu}u\!\cdot\!\nabla\beta
\!+\!s_{\mu}s\!\cdot\!\nabla\beta\!=\!0\\
&u\!\cdot\!\nabla\ln{|2q\phi|^{2}}\!=\!0\\
\nonumber
&\nabla^{2}(u\!\cdot\!\nabla\beta)\!-\!2m\cos{\beta}s\!\cdot\!\nabla\beta u\!\cdot\!\nabla\beta-\\
&-2m\sin{\beta}s\!\cdot\!\nabla(u\!\cdot\!\nabla\beta)\!=\!0\\
&s_{\rho}u_{\sigma}\varepsilon^{\mu\rho\sigma\nu}\nabla_{\mu}\nabla^{2}\ln{|2q\phi|^{2}}\!=\!0
\end{eqnarray}
with the field equations 
\begin{eqnarray}
&s\!\cdot\!\nabla\ln{|2q\phi|^{2}}\!-\!2m\sin{\beta}\!=\!0\\
\nonumber
&\nabla^{2}(s\!\cdot\!\nabla\beta)\!-\!2m\cos{\beta}|s\!\cdot\!\nabla\beta|^{2}-\\
&-2m\sin{\beta}s\!\cdot\!\nabla (s\!\cdot\!\nabla\beta)\!-\!|2q\phi|^{2}\!=\!0\label{module}
\end{eqnarray}
and considering that when two vector equations are taken in scalar product with $s$ and $u$ they develop no constraint, then these amount to $8$ expressions as it should be in order to determine $4$ derivatives for the $2$ physical fields given by the module and the Yvon-Takabayashi angle.

These are the field equations we will consider.
%%%%%%%%%%%%%%%%%%%%%%%%%%%%%%%%%%%%%%%%%%%%%%%%%%%%%%%%%%%%%%%%%%%%%%%%%%%%%%%%%%%%%%%%%%%%%%%%%%%
%%%%%%%%%%%%%%%%%%%%%%%%%%%%%%%%%%%%%%%%%%%%%%%%%%%%%%%%%%%%%%%%%%%%%%%%%%%%%%%%%%%%%%%%%%%%%%%%%%%
\section{Special Frames}
In our development of the electrodynamics of one single spinor field we have simply implemented the assumptions of negligible gravity and no torsion, and now is the time to implement also the hypotheses of working in the frame that is at rest and with spin aligned along the third of the axes: the rest-frame is the one where the velocity of the spinor field is zero so that (\ref{u}) loses its spatial component and (\ref{s}) loses its time component; then the spin-eigenstate is the one for which the spin is aligned along the third of the axes so that (\ref{s}) has only its third component.

These two hypotheses are thus $u^{0}\!=\!1$ and $s^{3}\!=\!1$ as the only non-zero components of the velocity vector and the spin axial vector, and in this case we obtain the form
\begin{eqnarray}
&\nabla_{1}\beta\!=\!\nabla_{2}\beta\!=\!0\\
&\nabla_{0}\ln{|2q\phi|^{2}}\!=\!0\\
\nonumber
&\nabla^{2}(\nabla_{0}\beta)\!-\!2m\cos{\beta}s\!\cdot\!\nabla\beta\nabla_{0}\beta-\\
&-2m\sin{\beta}s\!\cdot\!\nabla(\nabla_{0}\beta)\!=\!0\\
&\nabla^{2}\nabla_{1}\ln{|2q\phi|^{2}}\!=\!\nabla^{2}\nabla_{2}\ln{|2q\phi|^{2}}\!=\!0\\
&\nabla_{3}\ln{|2q\phi|^{2}}\!-\!2m\sin{\beta}\!=\!0\\
\nonumber
&\nabla^{2}(\nabla_{3}\beta)\!-\!2m\cos{\beta}s\!\cdot\!\nabla\beta\nabla_{3}\beta-\\
&-2m\sin{\beta}s\!\cdot\!\nabla (\nabla_{3}\beta)\!-\!|2q\phi|^{2}\!=\!0
\end{eqnarray}
which is a form that is now perfectly suited to be written in a specific system of coordinates such as the Cartesian.

In coordinates of Cartesian type, they will immediately give that $\beta\!=\!\beta(t,z)$ and $\phi\!=\!\phi(x,y,z)$ with the remaining field equations being given according to the expressions 
\begin{eqnarray}
&\partial_{x}\nabla^{2}\ln{|2q\phi|^{2}}\!=\!\partial_{y}\nabla^{2}\ln{|2q\phi|^{2}}\!=\!0\\
&\partial_{t}(\nabla^{2}\beta\!+\!2m\partial_{z}\cos{\beta})\!=\!0\label{a}\\
&\partial_{z}(\nabla^{2}\beta\!+\!2m\partial_{z}\cos{\beta})\!-\!|2q\phi|^{2}\!=\!0\label{b}\\
&\partial_{z}\ln{|2q\phi|^{2}}\!-\!2m\sin{\beta}\!=\!0\label{c}
\end{eqnarray}
from which more information can be extracted: we have that from (\ref{a}) it is $\nabla^{2}\beta\!+\!2m\partial_{z}\cos{\beta}=F(z)$ and thus from (\ref{b}, \ref{c}) it is $|2q\phi|^{2}\!=\!F'$ and $2m\sin{\beta}\!=\!(\ln{F'})'$ with the consequence that $\phi\!=\!\phi(z)$ and $\beta\!=\!\beta(z)$ in general, so the remaining equations are given according to the form
\begin{eqnarray}
&(-\beta'\!+\!2m\cos{\beta})''\!=\!|2q\phi|^{2}\\
&(\ln{|2q\phi|^{2}})'\!=\!2m\sin{\beta}
\end{eqnarray}
which should now be solved exactly.

Plugging the first into the second gives an equation in which only the YT angle is present, and thus we may use the first of the above equations
\begin{eqnarray}
&(-\beta'\!+\!2m\cos{\beta})''\!=\!|2q\phi|^{2}
\end{eqnarray}
to obtain the module once the YT angle is known, while their combination given by the expression
\begin{eqnarray}
&(-\beta'\!+\!2m\cos{\beta})'''=\!2m\sin{\beta}(-\beta'\!+\!2m\cos{\beta})''
\end{eqnarray}
will be used to obtain the YT angle in general.

Calling $2mz\!=\!w$ and $\beta/2\!=\!\arctan{G}$ would allow us to simplify this equation down to the easier rational form
\begin{eqnarray}
\left(\frac{1-2G'-G^{2}}{1+G^{2}}\right)'''\!\!\!\!
\!=\!\frac{2G}{1+G^{2}}\left(\frac{1-2G'-G^{2}}{1+G^{2}}\right)''
\end{eqnarray}
and by computing the derivatives we can see that standard QED solutions verifying the constraint
\begin{eqnarray}
G(7\!-\!9G^{2})(1\!-\!G')\!=\!0
\end{eqnarray}
are admissible. However, they would give either
\begin{eqnarray}
G\!=\!a
\end{eqnarray}
or
\begin{eqnarray}
G\!=\!w\!+\!a
\end{eqnarray}
and that is respectively either
\begin{eqnarray}
\beta\!=\!2\arctan{a}
\end{eqnarray}
or
\begin{eqnarray}
\beta\!=\!2\arctan{(2mz\!+\!a)}
\end{eqnarray}
where $a$ is a generic integration constant, but by inserting these special solutions for the YT angle into the equation for the module we get the final constraint
\begin{eqnarray}
&q\phi\!=\!0
\end{eqnarray}
eventually implying that there is either no spinorial field or that there is no mutual interaction between the spinor and its own electrodynamic field in standard QED.

Quite clearly, the consequence of this is that either, on the one hand, we have only the electrodynamic field with a free propagation, or, on the other hand, we have both the electrodynamic field and the spinor field but they can have no interaction in any way whatsoever.
%%%%%%%%%%%%%%%%%%%%%%%%%%%%%%%%%%%%%%%%%%%%%%%%%%%%%%%%%%%%%%%%%%%%%%%%%%%%%%%%%%%%%%%%%%%%%%%%%%%
%%%%%%%%%%%%%%%%%%%%%%%%%%%%%%%%%%%%%%%%%%%%%%%%%%%%%%%%%%%%%%%%%%%%%%%%%%%%%%%%%%%%%%%%%%%%%%%%%%%
\section{Comments}
Summarizing our results, we proved how under the assumptions $XW \!=\!0$ and $\partial\theta
\!-\!\Omega\!=\!0$ choosing the rest-frame and spin-eigenstate of the third axis gives $q\phi
\!=\!0$ identically: reading the underlying mathematics, we may state that in absence of torsion-gravity it is always possible to find a system of reference in which an electrodynamically self-interacting spinor field can not exist at all in QED.

In order to avoid such a conclusion one is compelled to reconsider the hypothesis that either for a particle in its rest-frame one cannot choose a spin-eigenstate or that a particle cannot be in a rest-frame, or the assumption for which either we cannot neglect gravity or we cannot have a vanishing torsion. No matter where we look, all appear to be quite reasonable restrictions, but still, together they give field equations with no QED-like solutions and thus at least one of them \emph{must} clearly be logically wrong.

What we believe to be a possible answer is that despite being always possible to boost into the rest-frame and in it rotate into the spin-eigenstate, nevertheless these local transformations come at the cost of non-zero components in the spin connection: for example, if the matter distribution was to display a spin precession, then it may be that in the rotating frame following the precession some non-inertial acceleration arises for which the spin connection is no longer zero, despite still having no curvature.

What this implies is that even in absence of gravity we should have $\partial\theta
\!-\!\Omega\!\neq\!0$ and because such a quantity does transform covariantly for a local change of frame then it is non-zero in every possible frame of reference.

Because non-inertial effects are those removable via a choice of frame then they cannot be encoded in an object that cannot ever vanish: we know of no non-gravitational physical field that can be contained in such an object.

In absence of this non-inertial yet never-vanishing effect, we might be confident that $\partial\theta\!-\!\Omega\!\neq\!0$ if and only if gravity is present, and therefore the contradiction we got was due to our neglecting the gravitational field.

If gravity really is negligible in this case, then the way out may be given by the torsional interaction.
%%%%%%%%%%%%%%%%%%%%%%%%%%%%%%%%%%%%%%%%%%%%%%%%%%%%%%%%%%%%%%%%%%%%%%%%%%%%%%%%%%%%%%%%%%%%%%%%%%%
%%%%%%%%%%%%%%%%%%%%%%%%%%%%%%%%%%%%%%%%%%%%%%%%%%%%%%%%%%%%%%%%%%%%%%%%%%%%%%%%%%%%%%%%%%%%%%%%%%%
\section{Conclusion}
In this paper, we have considered the spinor field that is usually described in electrodynamics, and that is with neither gravity nor torsion and within the rest-frame and spin-eigenstate: in terms of a mathematical analysis, we have proven that the system of the field equations cannot possess any solution. We have discussed how this might require revising the idea of rest-frame or spin-eigenstate, albeit it is more reasonable to assume that problems are due to neglecting the torsion or the gravitational field.

The renormalization of general quantum field theories in a gravitational background with torsion requires the interaction of torsion to be non-minimal with scalar and spinor fields: and in particular, spinor fields and torsion interact in a non-minimal way. This is required for quantum field theories to be consistent, since the interactions between second-quantized fields yields divergences, and consequently renormalization procedures must be set in.

The renormalization protocols in quantum field theories are well defined only when curvature and torsion are taken together in the same framework. Within the set-up of second-quantization this has been established, and for the present article we aimed to straightforward reasoning that could encompass also first-quantized fields as well.

Here we have argued that torsion and curvature should be considered already in first-quantized formalism.
%%%%%%%%%%%%%%%%%%%%%%%%%%%%%%%%%%%%%%%%%%%%%%%%%%%%%%%%%%%%%%%%%%%%%%%%%%%%%%%%%%%%%%%%%%%%%%%%%%%
\begin{acknowledgments}
Dr. Rold\~{a}o da Rocha is grateful to CNPq (Grant No. 303293/2015-2) and to FAPESP (Grant No.~2017/18897-8) for partial financial support.
\end{acknowledgments}
%%%%%%%%%%%%%%%%%%%%%%%%%%%%%%%%%%%%%%%%%%%%%%%%%%%%%%%%%%%%%%%%%%%%%%%%%%%%%%%%%%%%%%%%%%%%%%%%%%%
%%%%%%%%%%%%%%%%%%%%%%%%%%%%%%%%%%%%%%%%%%%%%%%%%%%%%%%%%%%%%%%%%%%%%%%%%%%%%%%%%%%%%%%%%%%%%%%%%%%

%%%%%%%%%%%%%%%%%%%%%%%%%%%%%%%%%%%%%%%%%%%%%%%%%%%%%%%%%%%%%%%%%%%%%%%%%%%%%%%%%%%%%%%%%%%%%%%%%%%
\end{document}